\documentclass[preprint,preprintnumbers,amsmath,amssymb]{revtex4}

\usepackage{graphicx}
\usepackage{dcolumn}
\usepackage{bm}

\begin{document}

\title{ The wave functions in the presence of constraints - Persistent Current in Coupled Rings }

\author{David Schmeltzer }

\affiliation{Physics Department, City College of the City University of New York 
New York, New York 10031} 




\begin{abstract}

We  present a new method for computing the wave function in the presence of constraints.  
As an explicit example we compute the wave function  for the many electrons   problem  in  coupled  metallic rings  in the presence of  external magnetic fluxes.  For equal fluxes and an even number of electrons  the constraints  enforce a wave function with   a vanishing total momentum and a large  persistent current and magnetization in  contrast to the odd number of electrons where at finite temperatures the current is suppressed. We propose that the  even-odd property  can be verified  by measuring the magnetization as a function of a  varying gate voltage coupled to the rings. By reversing the flux in one of the ring the current and magnetization vanish in both rings; this can be used as a non-local control device.

\end{abstract}

\maketitle

\vspace{0.2 in}

 The electronic wave function  in quantum  nanosystems at low temperatures is sensitive to interactions and  topology such as  the  genus number $g$ [1,2]  (the number  of holes on a closed surface).  As a result, the wave function has to satisfy certain   $constraints$, which generate    conserved  currents  [3,4]. The implementation of the constraints is a non-trivial  task  in Quantum Mechanics [4]. The root of the  difficulty is that for a given constraint the  hermitian conjugate constraint  operator might not be a constraint, therefore a  reduction of the phase space is not possible [4] . This problem is solved by including non-physical $ghost$  fields  [4].  In Classical  Mechanics   second class  constraints [4] are  solved  by replacing the $Poisson$ brackets by the $Dirac$ $bracket$ and  quantization is performed according to the  $Dirac$ correspondence principle [4,5] with the unpleasant feature   that the quantum  representation for  the operators 
 might  not always be  possible.  Here we will solve the constraints  without  the  need to introduce non-physical operators. 

As a model problem we will consider the    Aharonov-Bohm  geometry [6-11] for the case that the genus  is $g=2$. This corresponds to a double   ring structure  perfectly glued at one point to form 
a character ``8'' structure (see Fig. 1).  Such a structure   gives rise to an interesting   Quantum Mechanical problem  [12]. $Gluing$ the  two rings at the common  point $x=0$ gives rise to a constraint problem, which was  solved numerically  using the  $Dirac$ $brackets$  [4,12].

In Sec. II  we present  the  proposed  method for computing the wave function with constraints. 
We will work with a  folded geometry, therefore the problem will be equivalent to a two-component $spinor$  on  a single ring. The constraint is such  that  at the common point $x=0,L$  the  $annihilation$ $operators$ are   identified as a single  operator  $ [C_{1}(x)-C_{2}(x)]|_{x=0,L}=0$. 

Using the $Dirac$ method [3] we compute  the $Noether$  currents [4]  which allow us to  identify the  constraint currents. In the presence of external fluxes the constraints are translated into a set  of equations for the wave function.   The constraints induce correlations between the different components of the wave function.  For  non-interacting electrons   the wave function for $N$ electrons  is given by  the $Slater$ $determinant$ of the single particle states, but the current is the same if we sum over the single electronic  states. For the present  problem we must work with the many-body  wave function  of the two rings (which is not a simple product of the two rings).

\vspace{0.1 in}

In Sec. III we discuss the constraint method with the scattering theory and show that the strong coupling limit between the rings corresponds to the constraint problem considered in Sec. I.
In  Sec. IV we use the constraint method to compute the many-body wave function for two rings in the presence of constraints. We find that the many particle wave function built from the single particle wave function which obeys the constraints  is different from the many-body wave function which obeys the constraints. We show that the constraints impose additional relations between the amplitudes of the many-body wave function. In Sec. V we present the modification needed in order to include the physical geometry of the rings, e.g. finite thickness. As a concrete example we choose two narrow cylinders which are in contact on the line $x=0$.  Section VI is devoted to discussions.

\vspace{0.2 in }

\section{The Constraint Method For Two rings}

\vspace{0.2 in }

We consider two rings threaded by a magnetic flux $\Phi_\alpha$, where $\alpha = 1, 2$ represent the index for  each ring  $\varphi_\alpha = 2 \pi (\frac{e \Phi_\alpha}{h c} ) = 2 \pi \frac{\Phi_\alpha}{\Phi_0} \equiv 2 \pi \hat{\varphi}_\alpha$. The  rings have  a common point at  $y=0$ (Fig. 1).  The first ring is  restricted to the region   $0\leq y \leq L$  with  the single particle creation and annihilation operator obeying  periodic boundary conditions   $C(y+L) = C(y)$  and $C^{\dagger}(y+L) = C^{\dagger}(y)$.
The second ring is restricted to  $-L\leq y \leq 0$  with similar boundary conditions  $C(y-L) = C(y)$  and $C^{\dagger}(y-L) = C^{\dagger}(y)$. We  introduce two sets of operators:   for the first ring $0\leq  y \leq L$,  $C_{1}(x)=C(x)=C(y)$, and  $C^{\dagger}_{1}(x)=C^{\dagger}(x)=C^{\dagger}(y)$ and for the second ring   $-L\leq y \leq 0$,   $C_{2}(x)=C(-x)=C(y)$, $C^{\dagger}_{2}(x)=C^{\dagger}(-x)=C^{\dagger}(y)$. Due to the folding, two equal  fluxes  $\hat{\varphi}_{1}=\hat{\varphi}_{2}\equiv \hat{\varphi}$ will be described by two  opposite fluxes.  
\begin{equation}
H=\int_0^L\,dx\left[\frac{\hbar^{2}}{2m}C_{1}^{\dagger}(x)\left(-i\partial_{x}-\frac{2\pi}{L}\hat{\varphi}_{1}\right)^2C_{1}(x)+\frac{\hbar^{2}}{2m}C_{2}^{\dagger}(x)\left(i\partial_{x}-\frac{2\pi}{L}\hat{\varphi}_{2}\right)^2C_{2}(x)\right]. 
\label{hamiltonian}
\end{equation}

\vspace{0.1 in}


\vspace{0.1 in}


 



\noindent\textbf{The Continuity Constraint}
 
The Hamiltonian given in Eq. 1 will be investigated under the condition that the annihilation operators at the contact point must be identified as one operator. This is implemented with the help of the   constraint operator: 

\begin{equation}
\eta\equiv  [C_{1}(x)-C_{2}(x)]|_{x=0,L};\hspace{0.2 in} \eta |\chi,N>=0
\label{con}
\end{equation}

Following Dirac [3], the constraints and the time derivative of the constraints must be satisfied at any time.  We must  have $\eta |\chi,N>=0$ and  $\frac{d}{dt}\eta|\chi,N>=0$. 

\noindent\textbf{The Eigenvalue  Constraint} 

In order to satisfy the constraint at any time we need to show that  $\frac{d}{dt}\eta|\chi,N>=0$. This new equation is determined by  the time evolution of the constraint operator $\eta$. Following Dirac [3]  we introduce a Lagrange   multiplier $\lambda$   and replace the Hamiltonian $H$ by the total Hamiltonian  [3]: $H_{t}=H+\lambda \eta^{\dagger}\eta $, 
where $\eta$ is the constraint and $\lambda$ is   the  Lagrange multiplier. 
Using the Heisenberg equation of motion we obtain:

$\frac{d}{dt}\eta|\chi,N>=\frac{1}{i\hbar}[\eta,H_{t}]|\chi,N>= \frac{1}{i\hbar}([\eta,H]+ \lambda \eta)|\chi,N>= \frac{1}{i\hbar}[\eta,H]|\chi,N>0$.  

\noindent In obtaining this result we have used the relations :
$[\eta,\lambda \eta^{\dagger}\eta]|\chi,N>= \lambda \eta|\chi,N>$. Using the constraint condition $\eta |\chi,N>=0$ we find that  the condition $\frac{d}{dt}\eta|\chi,N>=0$ generates a new constraint operator which we identify as the   $eigenvalue$ constraint   operator $E$  given by   $[\eta,H]\equiv\frac{\hbar^2}{2m}E$:

\begin{equation}
E\equiv\left[\left(-i\partial_{x}-\frac{2\pi}{L}\hat{\varphi}_{1}\right)^2C_{1}(x)-\left(-i\partial_{x}+\frac{2\pi}{L}\hat{\varphi}_{2}\right)^2C_{2}(x)\right]\bigg{|}_{x=0,L}\hspace{0.1in};\hspace{0.2in} E|\chi,N>=0. 
\label{energy}
\end{equation}

\noindent\textbf{The Current  Constraint} 

The current constraint is obtained from the $Noether$ current. We perform an infinitesimal gauge transformation  $\widetilde{C}^{\dagger}(x)|0>= e^{i\epsilon(x)}C^{\dagger}(x)$. Due to the folding this transformation translates into: $\widetilde{C}^{\dagger}_{1}(x)|0>= e^{i\epsilon(x)}C^{\dagger}_{1}(x)|0>$ and  $\widetilde{C}^{\dagger}_{2}(x)|0>=e^{i\epsilon(-x)}C^{\dagger}_{2}(x)|0>$. The state $|\chi,N>$  must be  invariant  under a periodic gauge transformation $\epsilon(x)= \epsilon(x+L)$. 
  As a result of the transformation the    Hamiltonian $h=\frac{\hbar^{2}}{2m}[\delta_{\alpha,1}(-i\partial_{x}-\frac{2\pi}{L}\hat{\varphi}_{1})^2+\delta_{\alpha,2}(-i\partial_{x}+\frac{2\pi}{L}\hat{\varphi}_{2})^2]$ \hspace{0.1 in} is replaced by     $\widetilde{h}\equiv
\frac{\hbar^{2}}{2m}[\delta_{\alpha,1}(-i\partial_{x}-\frac{2\pi}{L}\hat{\varphi}_{1} +\partial_{x}(\epsilon(x)))^2+\delta_{\alpha,2}(-i\partial_{x}+\frac{2\pi}{L}\hat{\varphi}_{2}+\partial_{-x}(\epsilon(-x)))^2]$. The constraint is invariant under the gauge transformation  $\eta^{\dagger}(x)\eta(x)=\widetilde{\eta}^{\dagger}(x)\widetilde{\eta}(x)$. The constraint operator $\eta$ is replaced by the transformed one  $\widetilde{\eta}\equiv [e^{-i\epsilon(x)}\eta(x)]|_{x=0,L}\equiv[e^{-i\epsilon(x)}\widetilde{C}_{1}(x)-e^{-i\epsilon (-x)}\widetilde{C}_{2}(x)]|_{x=0,L}$ , $\widetilde{\eta}|\chi,N>=0$.   [$\epsilon(x)$ is an arbitrary periodic function in $L$, which is  continuous at $x=0$  and  has  a continuous derivative $\partial_{x}(\epsilon(x))\neq 0$  at $x=0$. For example, any function with the Fourier expansion  $\epsilon(x)=\sum_{r=1}^{r=\infty}\hat{\epsilon} _{r}sin[\frac{2\pi r}{L} x]$ and  Fourier components  $\sum_{r=1}^{r=\infty}\hat{\epsilon} _{r}\neq 0$
obeys this condition.] 
The transformed constraint  $\widetilde{\eta}|\chi,N>=0$ must hold at any time, therefore we have the equation : $\frac{d}{dt}\widetilde{\eta}|\chi,N>=0$. Applying the Heisenberg equation of motion for the transformed Hamiltonian   $\widetilde{h}$  and keeping only  first order terms in  $\partial_{x}(\epsilon(x))$  that obey $\partial_{x}(\epsilon(x))|_{x=0}\neq 0$ gives us:
\begin{eqnarray} &&i\hbar\frac{d}{dt}\widetilde{\eta}|\chi,N>=\frac{\hbar^{2}}{2m}\int_{0}^{L}\,dx[\widetilde{\eta},\widetilde{C}^{\dagger}_{1}(x)(-i\partial_{x}-\frac{2\pi}{L}\hat{\varphi}_{1}+\partial_{x}(\epsilon(x)))^2\widetilde{C}_{1}(x)\nonumber\\&&+\widetilde{C}^{\dagger}_{2}(x)(-i\partial_{x}+\frac{2\pi}{L}\hat{\varphi}_{2}+\partial_{-x}(\epsilon(-x)))^2\widetilde{C}_{2}(x)]|\chi,N>=0.
\label{Noether}
\end{eqnarray} 
Using  the energy constraint $E|\chi,N>=0$  we identify   the  $current$  continuity  constraint $\beta$: 
\begin {equation}
\beta=  \left[\left(-i\partial_{x}-\frac{2\pi}{L}\hat{\varphi}_{1}\right)C_{1}(x)+ \left(-i\partial_{x}+\frac{2\pi}{L}\hat{\varphi}_{2}\right)C_{2}(x)\right]|_{x=0,L}\hspace{0.1 in};\hspace{0.2 in} \beta|\chi,N>=0. 
\label{newconstraints}
\end{equation}

 \textit{ To conclude the  eigenstate  $|\chi,N>$  for $N$ particles in two rings must satisfy the following equations}:
\textbf{\begin{equation}
H|\chi,N>=E(N)|\chi,N>\hspace{0.1 in};\hspace{0.1 in} \eta|\chi,N>=0 \hspace{0.1 in}; \hspace{0.1 in}E|\chi,N>=0\hspace{0.1 in};\hspace{0.1 in}\beta|\chi,N>=0
\label{eigenvalue}
\end{equation}}

The eigenfunctions will be given in terms of the amplitude wave functions:
For example  the single particle state is given by : 

$|\chi,N=1>=\int_{0}^{L}\,dx[f_{1}(x)C^{\dagger}_{1}(x)+f_{2}(x)C^{\dagger}_{2}(x)]|0>.$ 

Similarly the two particle state is given by:

\begin{equation}
|\chi,N=2>=\int_{0}^{L}\,dx \int_{0}^{L}\,dy[f_{1,1}(x,y)C^{\dagger}_{1}(x)C^{\dagger}_{1}(y) +f_{1,2}(x,y)C^{\dagger}_{1}(x)C^{\dagger}_{2}(y) + f_{2,2}(x,y)C^{\dagger}_{2}(x)C^{\dagger}_{2}(y)]|0>]. 
\label{twop}
\end{equation}

The amplitudes  $f_{1}(x)$,$f_{2}(x)$ and  $f_{1,1}(x,y)$, $f_{1,2}(x,y)$ $f_{2,2}(x,y)$ are determined by the condition given in Eq. (8).

\noindent\textbf{The Current Operator}

\vspace{0.1 in}

The $N$ particle wave function $< x_{N},...x_{1}|\chi,N>$ must obey   periodic boundary conditions:

 $<0|C_{\alpha_{1}}(x_{1})..  C_{\alpha_{k}}(x_{k})..C_{\alpha_{N}}(x_{N})|\chi,N>=<0|C_{\alpha_{1}}(x_{1})..  C_{\alpha_{k}}(x_{k}+L)..C_{\alpha_{N}}(x_{N})|\chi,N>$
 where $\alpha_{i}$ takes two values $\alpha_{i}=1$  or $\alpha_{i}=2$.
 
Once the eigenfunction $|\chi,N>$ is known we can use the     current operators  $\hat{J}_{1}(x)$ and $\hat{J}_{2}(x)$ in the second quantized form  to compute the current in   each ring:

\begin{eqnarray}
&&\hat{J}_{1}(x)=\frac{\hbar}{i 2m}[C^{\dagger}_{1}(x)(\partial_{x}-i\frac{2\pi}{L}\hat{\varphi}_{1})C_{1}(x)-((\partial_{x}-i\frac{2\pi}{L}\hat{\varphi}_{1})C^{\dagger}_{1}(x))C_{1}(x) \hspace{0.1 in}; J_{1}(x)=\frac{<N,\chi|\hat{J}_{1}(x)|\chi,N>}{<N,\chi|\chi,N>}, \nonumber\\&&
\hat{J}_{2}(x)=\frac{\hbar}{i2m}[C^{\dagger}_{2}(x)(\partial_{x}+i\frac{2\pi}{L}\hat{\varphi}_{2})C_{2}(x)-((\partial_{x}+i\frac{2\pi}{L}\hat{\varphi}_{2})C^{\dagger}_{2}(x))C_{2}(x)\hspace{0.1 in};  J_{2}(x)=\frac{<N,\chi|\hat{J}_{2}(x)|\chi,N>}{<N,\chi|\chi,N>}. 
\label{eqnarray}
\end{eqnarray}

\vspace{0.2 in}



\section{ The Emerging Constraint Conditions From The Tight Binding Formulation}

\vspace{0.1 in} 

The Hamiltonian in Eq. (1) must be supplemented by  the coupling term between the rings.
The most general form for the coupling is given by:

\begin{eqnarray} 
 H_{coupling}&=&\int_0^L\,dx \delta(x)[-U_{\bot}(C^{\dagger}_{1}(x)C_{2}(x)+C^{\dagger}_{2}(x)C_{1}(x))+U_{||}(C^{\dagger}_{1}(x)C_{1}(x)+C^{\dagger}_{2}(x)C_{2}(x))]\nonumber\\&&=\int_0^L\,dx \delta(x)[U_{\bot}( C^{\dagger}_{1}(x)-C^{\dagger}_{2}(x)) ( C_{1}(x)-C_{2}(x))+(U_{||}-U_{\bot})(C^{\dagger}_{1}(x)C_{1}(x) \nonumber\\ 
 &&+C^{\dagger}_{2}(x)C_{2}(x))].
\label {eqnarray}  
\end{eqnarray}

\noindent We introduce the notation  $U_{\bot} =t U$ and   $U_{||}=s U$  where $s$ and $t$ are parameters. Using the spinor representation $ \hat{C}(x)=[C_{1}(x),C_{2}(x)]^{T}$  we can rewrite the coupling Hamiltonian in terms of the Pauli matrix $\sigma_{1}$ and the identity matrix  $I$: 

 $ H_{coupling}=\int_0^L\,dx\delta(x)U\Psi^{+} (x)[(s I- t\sigma_{1})]\Psi (x).$ 
 
 
 
\noindent This problem belongs to  the class of delta function  potential considered in  Quantum Mechanics.
 
\vspace{0.1 in}

\noindent\textbf{A-The wave function for a single particle, N=1}

 $|\chi,N=1>=\int_{0}^{L}\,dx[f_{1}(x)C^{\dagger}_{1}(x)+f_{2}(x)C^{\dagger}_{2}(x)]|0>$, 


$(H+ H_{coupling})|\chi,N=1>=E(1)|\chi,N=1>$. 

\noindent As a result we obtain the Schr\"odinger equation  in terms of the two amplitudes $ f_{1}(x)$,  $f_{2}(x)$. The Hamiltonian in Eq. (1)  together with   $H_{coupling}$ can be solved  using   the method for    delta function potentials.  We integrate the single particle Shr\"odinger equation  around $x=0,L$ and obtain  the discontinuity   $derivative$ of  the spinor  $\Psi (x)=[f_{1}(x),f_{2}(x)]^{T}$ which obeys  $ \frac{d \Psi (x)}{dx}|^{x=\epsilon}_{x=-\epsilon}\equiv \frac{d \Psi (x)}{dx}|^{x=\epsilon}_{x= L-\epsilon}$.  

\vspace{0.2 in}

$[(-i\partial_{x}-\frac{4\pi}{L}\hat{\varphi}_{1})f_{1}(x)|^{x=\epsilon}_{x=L-\epsilon}=\frac{-i 2m}{\hbar^2} U \frac{1}{2}[( s f_{1}(\epsilon)-  t f_{2}(\epsilon)) + (s f_{1}(L-\epsilon)- t f_{2}(L-\epsilon))]$,

\vspace{0.2 in}

$(-i\partial_{x}+\frac{4\pi}{L}\hat{\varphi}_{2})f_{2}(x)|^{x=\epsilon}_{x=L-\epsilon}=
\frac{-i 2m}{\hbar^2} U\frac{1}{2}[( s f_{2}(\epsilon)-  t f_{1}(\epsilon)) + (s f_{2}(L-\epsilon)- t f_{1}(L-\epsilon))]$. 

This set of equations gives us the boundary conditions for the present problem.  \textit{Indeed these equations are determined by the discontinuity   function $ U[s f_{2}(0)- t f_{1}(0)]$.} For this case the solution follows from the method  of the delta function potential--see Griffiths  Quantum Mechanics section, 2.5 page 73. 

\vspace{0.1 in}

\noindent\textbf{B-The wave function for two  particles, N=2}

\vspace{0.1 in}

In order to compute the wave function for $N$ particles  we have to compute the boundary conditions for the amplitudes of the wave function.
We will consider the case of two particles which can be generalized to many particles.
$|\chi,N=2>=\int_{0}^{L}\,dx_{1} \int_{0}^{L}\,dx_{2}[f_{1,1}(x_{1},x_{2})C^{\dagger}_{1}(x_{1})C^{\dagger}_{1}(x_{2}) +f_{1,2}(x_{1},x_{2})C^{\dagger}_{1}(x_{1})C^{\dagger}_{2}(x_{2}) + f_{2,2}(x_{1},x_{2})C^{\dagger}_{2}(x_{1})C^{\dagger}_{2}(x_{2})]|0>]$. 

\vspace{0.1 in}

Using the eigenvalue equation:  $(H+ H_{coupling})|\chi,N=2>=E(2)|\chi,N=2>$ we  
integrate the two  particle Shr\"odinger equation  around $x_{1}=0,L$ and obtain  the discontinuity   $derivative$  for the three amplitudes  $f_{1,1}(x_{1},x_{2})$, $f_{2,2}(x_{1},x_{2})$, $f_{1,2}(x_{1},x_{2})$

$[(-i\partial_{x_{1}}-\frac{4\pi}{L}\hat{\varphi}_{1})f_{1,1}(x_{1},x_{2})|^{x_{1}=\epsilon}_{x_{1}=L-\epsilon}=\frac{-i 2m}{\hbar^2} U \frac{1}{2}[ s (f_{1,1}(x_{1}=\epsilon,x_{2})+ f_{1,1}(x_{1}=L-\epsilon,x_{2}))]$, 

$[(-i\partial_{x_{1}}-\frac{4\pi}{L}\hat{\varphi}_{1})f_{2,2}(x_{1},x_{2})|^{x_{1}=\epsilon}_{x_{1}=L-\epsilon}=\frac{-i 2m}{\hbar^2} U \frac{1}{2}[- t (f_{2,2}(x_{1}=\epsilon,x_{2})+ f_{2,2}(x_{1}=L-\epsilon,x_{2}))]$, 

$[(-i\partial_{x_{1}}-\frac{4\pi}{L}\hat{\varphi}_{1})f_{1,2}(x,x_{2})|^{x_{1}=\epsilon}_{x_{1}=L-\epsilon}=\frac{-i 2m}{\hbar^2} U \frac{1}{2}[(s- t) (f_{1,2}(x_{1}=\epsilon,x_{2})+ f_{1,2}(x_{1}=L-\epsilon,x_{2}))]$. 

\noindent Similar equations are obtained by exchanging $x_{1}$ with $x_{2}$. This set of equations determines the two particle  wave function $<x_{1},x_{2}|\chi,N=2>$. This procedure is rather involved  but can be generalized to the $N$ particles case.

\noindent\textbf{C-The strong coupling limit $U\rightarrow\infty$ }

Next we investigate the strong coupling limit and show that the problem can be simplified to a constraint problem.  We consider the case  $s=t=1$ for which we have  the  scattering matrix $S$  given by:
 

$S=T \exp{^{-i\frac{U}{\hbar}\int_{-infty}^{infty}\,d\tau [(C^{+}_{1}(x=0,\tau)-C^{+}_{2}(x=0,\tau)) (C_{1}(x=0,\tau)-C_{2}(x=0,\tau))]}}$.   For  $U\rightarrow\infty$     the  scattering matrix $S$  obeys :
  
$ lim_{U\rightarrow\infty} T e^{-i\frac{U}{\hbar}\int_{-\infty}^{\infty}\,d\tau [(C^{+}_{1}(x=0,\tau)-C^{+}_{2}(x=0,\tau)) (C_{1}(x=0,\tau)-C_{2}(x=0,\tau))]}|\chi,N>\rightarrow(C_{1}(x=0,\tau)-C_{2}(x=0,\tau))|\chi,N>=0$.

\noindent As a result the  field  $(C_{1}(x=0,\tau)-C_{2}(x=0,\tau))$ is enforced to satisfy     $C_{1}(x=0,\tau)-C_{2}(x=0,\tau)=0 $  which is equivalent to the constraint condition:  
  
$\eta |\chi,N>\equiv  [C_{1}(x)-C_{2}(x)]|_{x=0,L}|\chi,N>=0$. 
  





\vspace{0.1 in}

 
\vspace{0.1 in}

\vspace{0.1 in}
 
\section{Computation of the Wave Function  for Equal Fluxes }

\vspace{0.1 in}

\textit{For the strong coupling limit we will use the constraints given by the equation 6.}
When the   fluxes  are  the same for both rings  the constraint operator $\beta$ is simplified to a new constraint   $\gamma=i\beta(\hat{\varphi}_{1}= \hat{\varphi}_{2})$:
\begin{equation}
\gamma=[\partial_{x}C_{1}(x)+\partial_{x}C_{2}(x)]|_{x=0,L}\hspace{0.1 in}; \hspace{0.1 in} \gamma|\chi,N>=0. 
\label{newconstraint}
\end{equation}
The  $N$ particles  wave function  for equal fluxes must satisfy the following   conditions :
\begin{equation}
H|\chi,N>=E(N)|\chi,N>\hspace{0.1 in};\hspace{0.1 in} \eta|\chi,N>=0 \hspace{0.1 in}; \hspace{0.1 in}E|\chi,N>=0\hspace{0.1 in};\hspace{0.1 in}\gamma|\chi,N>=0. 
\label{efluxcond}
\end{equation}

\vspace{0.1 in}

\noindent\textbf{A-The single particle case}

The   $single$ particle case  corresponds to one  electron in two rings. The state for one particle is given by:
  $|\chi,N=1>=\int_{0}^{L}\,dx[f_{1}(x)C^{\dagger}_{1}(x)+f_{2}(x)C^{\dagger}_{2}(x)]|0>$.  
The  two-component spinor  amplitudes  $ f_{1}(x)$ and $f_{2}(x)$  represent the wave  function. Using the Hamiltonian given in Eq. (1) we can write down  the eigenvalue equation $H|\chi,N=1>=E(1)|\chi,N=1>$. A standard calculation shows  this equation is equivalent to two  eigenvalue equations for the amplitudes    $ f_{1}(x)$ and $f_{2}(x)$.
\begin{equation}
\frac{\hbar^{2}}{2m}\left(-i\partial_{x}-\frac{2\pi}{L}\hat{\varphi}\right)^2 f_{1}(x)=E(1)f_{1}(x) 
\hspace{0.1 in};\hspace{0.1 in}
\frac{\hbar^{2}}{2m}\left(-i\partial_{x}-\frac{2\pi}{L}\hat{\varphi}\right)^2 f_{2}(x)=E(1)f_{2}(x).
\label{eigen}
\end{equation}
 The constraint operators   given in Eq. (8) generate  the followings boundary  conditions at $x=0 $:
\begin{equation}
f_{1}(x=0)=f_{2}(x=0) ;\hspace{0.1 in}[\partial_{x}f_{1}(x)+\partial_{x}f_{2}(x)]|_{x=0}=0. \label{wavefunction}
\end{equation} 
The first equation  is equivalent to the continuity of the wave function at $x=0$ and the  second equation describes the continuity of the derivative of the  wave function (once we fold back the space ) at $x=0$.
From the eigenvalue equation given in Eq. (9) we find:   $E(n;N=1)=\frac{\hbar^{2}}{2m}(\frac{2\pi}{L})^2 (n-\hat{\varphi})^2$ for the ring one  and  $E(-n;N=1)=\frac{\hbar^{2}}{2m}(\frac{2\pi}{L})^2 (n-\hat{\varphi})^2$  for the second ring. Due to the folding of the space around x=0, the eigenvalue with the quantum  number $n$ in ring one and Quantum number $-n $  in the second ring  are equal. This result holds for the quantum numbers,   $n=0,\pm1, \pm2,....$ . The single particle state   $|n,N=1>$ for  $\hat{\varphi} \neq \frac{1}{2}$ is given by:
\begin{equation}
|n;N=1>=\frac{1}{\sqrt{2L}}\int_{0}^{L}\,dx[e^{i\frac{2\pi}{L}n x}C^{\dagger}_{1}(x)+e^{-i\frac{2\pi}{L}n x}C^{\dagger}_{2}(x)]|0>. 
\label{n=1}
\end{equation}
To understand this result we fold back the ring such that $x\rightarrow-x$. This means that if the particle in the first ring ($x<0$) has the momentum $ \frac{2\pi}{L}n$ it will be perfectly transmitted to the second ring with the same momentum and the same amplitude.    
If we remove the point $x=0$ and create  a ring of a double length $2L$, the current will be the same as in one ring with the same flux. Indeed, the only difference being the doubling of the size. As a result, we will have half of the  current in a  single ring. (If we rescale the length, we find the same current as in one ring    [11].) It is important to remark  that the states $|n;N=1>$  and $|-n;N=1>$ correspond to two different eigenvalues. Therefore, for a given eigenvalue we can not have a linear combination of waves  $e^{i\frac{2\pi}{L}n x}$ and $e^{-i\frac{2\pi}{L}n x}$ in the same ring. The  wave $e^{i\frac{2\pi}{L}n x}$ in ring one will be transmitted into  the second  ring without any reflection, the form of the transmitted wave will be  $e^{-i\frac{2\pi}{L}n x}$  (in the unfolded  coordinates the form of the wave will be      $e^{i\frac{2\pi}{L}n y}$ in the second  ring   for $y<0$).
In Fig. 1 we show the current flow for  two rings with  equal fluxes in the unfolded geometry. The 
current vanishes if we have the opposite flux in the two rings, as depicted in Fig. 2. 
   
\textit{The case  $\hat{\varphi}=\frac{1}{2}$ deserves special consideration.}
The  eigenvalue operator $E$ has  two pairs of momentum with the same eigenvalue:  The first pair $n_{1}=n$ in the first ring and  $n_{2}=-n$  for the second ring and the second pair  $n'_{1}=-n+2\hat{\varphi}$ (ring one) and   $n'_{2}=n-2\hat{\varphi}$ (ring two). As a result we obtain two degenerate eigenstates  $|n;N=1,+>$ and $|n;N=1,->$ given by:
\begin{eqnarray}
&&|n;N=1,+>=\frac{1}{\sqrt{2L}}\int_{0}^{L}\,dx[e^{i\frac{2\pi}{L}n x}C^{\dagger}_{1}(x)+e^{-i\frac{2\pi}{L}n x}C^{\dagger}_{2}(x)]|0>\hspace{0.1 in};\nonumber\\&& 
|n;N=1,->=\frac{1}{\sqrt{2L}}\int_{0}^{L}\,dx[e^{-i\frac{2\pi}{L}(n-2\hat{\varphi}) x}C^{\dagger}_{1}(x)+e^{i\frac{2\pi}{L}(n-2\hat{\varphi}) x}C^{\dagger}_{2}(x)]|0>. 
\label{N=1;-}
\end{eqnarray}

As a result the current for the state  $|n;N=1,->$ will be opposite to the current for the state $ |n;N=1,+>$ 
Since the two eigenstates  $|n;N=1,+>$ and $|n;N=1,->$  are degenerate, the single particle state  will be given  by two linear combinations of the  eigenstates  $|n;N=1,+>$ and $|n;N=1,->$:
 $|\chi(n),\hat{\varphi}=\frac{1}{2};N=1>=\alpha_{+} |n;N=1,+>\pm \alpha_{-}|n;N=1,+>$ with the condition  $|\alpha_{+}|^2+|\alpha_{-}|^2=1$.
 For the special values   $|\alpha_{+}|^2$=$|\alpha_{-}|^2$  the current will  vanish.
 
 \vspace{0.1 in}
  
\noindent\textbf{B-The Two particles state}

\vspace{0.1 in}
 
We will construct the two particles state and will show that due to the constraints not all the antisymmetric combination of the single particles states which obey the constraints are allowed. Imposing the constraints on the two particles state  imposes further  restrictions.
The $two$ particles eigenstate is determined by the three  components  $f_{11}(x_{1},x_{2})$, $f_{12}(x_{1},x_{2})$ and $f_{22}(x_{1},x_{2})$ that obey the eigenvalue equations:

$\frac{\hbar^{2}}{2m}[(-i\partial_{x_{1}}-\frac{2\pi}{L}\hat{\varphi})^2+ (-i\partial_{x_{2}}-\frac{2\pi}{L}\hat{\varphi})^2]f_{11}(x_{1},x_{2})=E(2)f_{11}(x_{1},x_{2})$,

$\frac{\hbar^{2}}{2m}[(-i\partial_{x_{1}}-\frac{2\pi}{L}\hat{\varphi})^2+ (i\partial_{x_{2}}-\frac{2\pi}{L}\hat{\varphi})^2]f_{12}(x_{1},x_{2})=E(2)f_{12}(x_{1},x_{2})$,

$\frac{\hbar^{2}}{2m}[(i\partial_{x_{1}}-\frac{2\pi}{L}\hat{\varphi})^2+ (i\partial_{x_{2}}-\frac{2\pi}{L}\hat{\varphi})^2]f_{22}(x_{1},x_{2})=E(2)f_{22}(x_{1},x_{2})$.

The amplitudes  $f_{11}(x_{1},x_{2})$,  $f_{12}(x_{1},x_{2})$ and $f_{22}(x_{1},x_{2})$ are constructed from  the single particle states which are represented in terms of the complex coordinate  $Z(x)=e^{i\frac{2\pi}{L} x}$ and  $Z^{*}(x)=e^{-i\frac{2\pi}{L} x}$. We  introduce the $antisymmetry$ $operator$   $\mathcal{\tilde A}$,  which acts both on the space coordinates and the  ring index matrices    $A_{11}$ (two particles on ring one), $ A_{12}$ (one particle on ring one and the second on ring two), and $A_{22}$ (two particles on ring two). When the operator $\mathcal{\tilde A}$ acts on a two particle wave function it gives :
$\mathcal{\tilde A}[A_{12}(Z(x_{1}))^{n}(Z(x_{2}))^{m}]\equiv [A_{12}Z(x_{1}))^{m}(Z(x_{2}))^{n}-A_{21}Z(x_{2}))^{n}(Z(x_{1}))^{m}]$  and $\mathcal{\tilde A}[A_{ii}(Z(x_{1}))^{n}(Z(x_{2}))^{m}]\equiv [A_{ii}Z(x_{1}))^{m}(Z(x_{2}))^{n}-A_{ii}Z(x_{2}))^{n}(Z(x_{2}))^{m}]$ 
for $i=1,2$.

From the eigenvalue constraint  $E|n,m;N=2>=0 $  we obtain the condition for the eigenvalues. The only possible solution for these equations are states  with $m=-n$  which give    eigenvalues $E(2)=E(n,-n;N=2)=\frac{\hbar^{2}}{2m}(\frac{2\pi}{L})^2[ (n-\hat{\varphi})^2+(-n-\hat{\varphi})^2]$,  $n=0,\pm1,\pm2..$.

For amplitude $f_{11}(x_{1},x_{2})$ we consider only   the single particle states with $n$ and $-n$ which have the  eigenvalue $E(2)= \frac{\hbar^{2}}{2m}(\frac{2\pi}{L})^2[ (n-\hat{\varphi})^2+(-n-\hat{\varphi})^2]$. We construct the antisymmetric  amplitudes will be given by:

 $f_{11}(x_{1},x_{2})=A_{11}[(Z(x_{1}))^{n}(Z(x_{2}))^{-n} -(Z(x_{2}))^{n}(Z(x_{1}))^{-n}]$. 
 
\noindent Similarly for two electrons on the second ring  $f_{22}(x_{1},x_{2})$  we have:

 $f_{22}(x_{1},x_{2})=B_{11}[(Z(x_{1}))^{-n}(Z(x_{2}))^{n} -(Z(x_{2}))^{-n}(Z(x_{1}))^{n}]$. 
 
\noindent The  amplitude for one electron on ring one and the second electron on ring two is given by  $f_{12}(x_{1},x_{2})$ .This corresponds to two pairs of states   $n$, $n$ and  $-n$ , $-n$.  The eigenvalue for the pair   $n$ , $n$  is equal to  $E(2)= \frac{\hbar^{2}}{2m}(\frac{2\pi}{L})^2[ (n-\hat{\varphi})^2+(-n-\hat{\varphi})^2]$. For  $-n$ , $-n$  we have the same eigenvalue.
The amplitude  $f_{12}(x_{1},x_{2})$ is given by the linear combination of the two pairs . Using the $antisymmetry$ $operator$   $\mathcal{\tilde A}$ we obtain the amplitude  $f_{12}(x_{1},x_{2})$   for the two pairs:

\begin{eqnarray*}
f_{12}(x_{1},x_{2})&=& \mathcal{\tilde A}[A_{12}(Z(x_{1}))^{n}(Z(x_{2}))^{n}]+\mathcal{\tilde A}[B_{12}(Z(x_{1}))^{-n}(Z(x_{2}))^{-n}]\nonumber\\&&= [A_{12}(Z(x_{1}))^{n}(Z(x_{2}))^{n}-A_{21}(Z(x_{2}))^{n}(Z(x_{1}))^{n}]+\nonumber\\&& [B_{12}(Z(x_{1}))^{-n}(Z(x_{2}))^{-n}-B_{21}(Z(x_{2}))^{-n}(Z(x_{1}))^{-n}].  \nonumber\\&&
\end{eqnarray*}

 Using constraints given in Eq. (8)  for the two particles state  $|n,m;N=2>$:
 $\eta|n,m;N=2>=0 $, $E|n,m;N=2>=0 $ and  $\gamma |n,m;N=2>=0$, and we obtain the following boundary conditions:
 
 $2f_{11}(x_{1},0)=f_{1,2}(x_{1},0)\hspace{0.15 in};\hspace{0.2 in} [2\partial_{x_{2}}f_{11}(x_{1},x_{2})+\partial_{x_{2}}f_{12}(x_{1},x_{2})]_{x_{2}=0}=0$,

$2f_{2,2}(x_{1},0)=f_{1,2}(0,x_{1})\hspace{0.1 in}; \hspace{0.1 in} [2\partial_{x_{2}}f_{22}(x_{2},x_{1})+\partial_{x_{2}}f_{12}(x_{1},x_{2})]_{x_{2}=0}=0$.

From these equations we find that the amplitudes obey the relations: $A_{12}=-A_{21}=2A_{11}$; $A_{11}=A_{22}$ and $B_{21}=-B_{12}=2A_{22}$. We introduce   the  antisymmetric  spinor  notation  $\epsilon_{1;2} \equiv\frac{A_{12}}{2}$,  which obeys the relations: $\epsilon^{1,1}_{1;2}=- \epsilon^{1,1}_{2;1}$  and  $(\epsilon^{1,1}_{1;2})^{\dagger}\cdot\epsilon^{1,1}_{1;2}=1$  (the upper index $1,1$  means that we have one electron in each ring, the bottom index  $1,2$ or $2,1$ represents the order . $\epsilon^{1,1}_{1;2}$  the first electron is  ring one and the second  electron is on ring two and $\epsilon^{1,1}_{2;1}$ represents the first electron on ring two and second electron on ring one.) 
 The normalized two particle state   is given by:
\begin{eqnarray}
&&|n,-n;N=2>=\nonumber\\&&
\int_{0}^{L}\,dx_{1} \int_{0}^{L}\,dx_{2}[f_{11}(x_{1},x_{2})C^{\dagger}_{1}(x_{1})C^{\dagger}_{1}(x_{2}) +f_{12}(x_{1},x_{2})C^{\dagger}_{1}(x_{1})C^{\dagger}_{2}(x_{2}) + f_{22}(x_{1},x_{2})C^{\dagger}_{2}(x_{1})C^{\dagger}_{2}(x_{2})]|0>\nonumber\\&&=\int_{0}^{L}\,dx_{1} \int_{0}^{L}\,dx_{2}\frac{1}{4 L}[[(Z(x_{1}))^{n}(Z^{*}(x_{2}))^{n}-(Z(x_{2}))^{n}(Z^{*}(x_{1}))^{n}]C^{\dagger}_{1}(x_{1})C^{\dagger}_{1}(x_{2})\nonumber\\&& +2\epsilon^{1,1}_{1;2}
[Z(x_{1}))^{n}(Z(x_{2}))^{n}-(Z^{*}(x_{2}))^{n}(Z^{*}(x_{1}))^{n}]C^{\dagger}_{1}(x_{1})C^{\dagger}_{2}(x_{2})\nonumber\\&& +[(Z^{*}(x_{1}))^{n}(Z(x_{2}))^{n}-(Z^{*}(x_{2}))^{n}(Z(x_{1}))^{n}]C^{\dagger}_{2}(x_{1})C^{\dagger}_{2}(x_{2})]|0>]. 
\label{twop}
\end{eqnarray}
The off-diagonal spinor component $f_{12}(x_{1},x_{2})\propto4isin(\frac{2\pi}{L}n((x_{1}+x_{2}))$ is symmetric in space and resembles  the $BCS$  pairing wave function (once we identify the ring index with the spin) in contrast to  the diagonal elements $f_{11}((x_{1},x_{2})$ and $f_{22}((x_{1},x_{2})$, which are antisymmetric in space. 
The two particles state, which obeys the constraints are  different from the two particles state constructed from the single particles, which obey the constraints.  Using the  single particle states $|n;N=1>$  and $|m;N=1>$ [which  obey Eq. (11)] we  construct  an antisymmetric tensor product  $|n,m;N=2>_{build}=|n;N=1>|m;N=1>-|m;N=1>|n;N=1>$.  This state  is not a solution which obeys the constraints for the two particles state. The only possibility is to have   an antisymmetric tensor product of two states with vanishing total momentum  $|n,-n;N=2>=|n;N=1>|-n;N=1>-|-n;N=1>|n;N=1>$.
(The ground state for  the two  particles ($\hat{\varphi}<\frac{1}{2}$) is given by the eigenstate   $|1,-1;N=2>$ .)
\textit{This structure persists for an  even  numbers of electrons  $N=2M$ and gives rise to a robust state  absent for the  single ring.} 

\vspace{0.1 in}

\noindent\textbf{C-The three particles state}

\vspace{0.1 in}

The  wave-function for $three$ particles can  only be found for special configurations  $|m,n,-n;N=3> $  $m\neq n$ and $m\neq -n$.
The ground state will be given by the state $|0,1,-1;N=3>$.   
The three particles  state is   determined by the four amplitudes  $f_{111}(x_{1},x_{2},x_{3})$,  $f_{112}(x_{1},x_{2},x_{3})$,  $f_{122}(x_{1},x_{2},x_{3})$ and  $f_{122}(x_{1},x_{2},x_{3})$, which obey the eigenvalue equation:

 $\frac{\hbar^{2}}{2m}[(-i\partial_{x_{1}}-\frac{2\pi}{L}\hat{\varphi})^2+ (-i\partial_{x_{2}}-\frac{2\pi}{L}\hat{\varphi})^2+(-i\partial_{x_{3}}-\frac{2\pi}{L}\hat{\varphi})^2]f_{111}(x_{1},x_{2},x_{3})=E(3)f_{111}(x_{1},x_{2},x_{3})$,

$\frac{\hbar^{2}}{2m}[(-i\partial_{x_{1}}-\frac{2\pi}{L}\hat{\varphi})^2+ (-i\partial_{x_{2}}-\frac{2\pi}{L}\hat{\varphi})^2+(-i\partial_{x_{2}}+\frac{2\pi}{L}\hat{\varphi})^2]f_{112}(x_{1},x_{2},x_{3})=E(3)f_{112}(x_{1},x_{2},x_{3})$,

$\frac{\hbar^{2}}{2m}[(-i\partial_{x_{1}}-\frac{2\pi}{L}\hat{\varphi})^2+ (-i\partial_{x_{2}}+\frac{2\pi}{L}\hat{\varphi})^2+(-i\partial_{x_{3}}+\frac{2\pi}{L}\hat{\varphi})^2]f_{122}(x_{1},x_{2},x_{3})=E(3)f_{122}(x_{1},x_{2},x_{3})$, 

$\frac{\hbar^{2}}{2m}[(-i\partial_{x}+\frac{2\pi}{L}\hat{\varphi})^2+ (-i\partial_{x_{2}}+\frac{2\pi}{L}\hat{\varphi})^2+(-i\partial_{x_{3}}+\frac{2\pi}{L}\hat{\varphi})^2]f_{222}(x_{1},x_{2},x_{3})=E(3)f_{222}(x_{1},x_{2},x_{3})$. 

\noindent Using Eq. (8) we obtain the following relations for the spinor components: 

$3f_{111}(x_{1},x_{2},0)=f_{112}(x_{1},x_{2},0) \hspace{0.1 in}; [3\partial_{x_{3}}f_{111}(x_{3},x_{1},x_{2})+\partial_{x_{3}}f_{112}(x_{2},x_{1},x_{3})]_{x_{3}=0}=0$,

$ 3f_{222}(0,x_{1},x_{2})=f_{122}(0,x_{1},x_{2}) \hspace{0.1 in}; [3\partial_{x{3}}f_{222}(x_{3},x_{1},x_{2})+\partial_{z}f_{112}(x_{2},x_{1},z_{3})]_{z_{3}=0}=0$,

 $2f_{121}(x_{1},x_{2},0)=f_{122}(x_{2},x_{1},0) \hspace{0.1 in};
[3\partial_{x_{3}}f_{121}(x_{1},x_{2},x_{3})+\partial_{x_{3}}f_{122}(x_{1},x_{2},x_{3})]_{z=0}=0.$

The solution of the constraint equations fixes  the eigenvalue and the state. The ground state  eigenvalue is given by  $E_{g}(0,1,-1;N=3)=\frac{\hbar^{2}}{2m}(\frac{2\pi}{L})^2 [(\hat{\varphi})^2+(1-\hat{\varphi})^2+(-1-\hat{\varphi})^2]$ and the three particles ground state is :
\begin{eqnarray}
&&|0,1,-1;N=3>=\nonumber\\&&\int_{0}^{L}\,dx_{1} \int_{0}^{L}\,dx_{2}\int_{0}^{L}\,dx_{3}[\Phi_{0,1,-1}(x_{1},x_{2},x_{3})C^{\dagger}_{1}(x_{2}C^{\dagger}_{1}(x_{2})C^{\dagger}_{1}(x_{3})\nonumber\\&& +3[\epsilon^{2,1}_{1,1;2}\sum_{i=x_{1},x_{2},x_{3}}\hat{P}_{i,x_{3}}(\Phi_{0,1}(x_{1},x_{2})Z(x_{3})-\Phi_{0,-1}(x_{1},x_{2})Z^{*}(x_{3}))+\Phi_{0,1,-1}(x_{1},x_{2},x_{3})]C^{\dagger}_{1}(x_{1})C^{\dagger}_{1}(x_{2})C^{\dagger}_{2}(x_{3})\nonumber\\&&+3[ \epsilon^{1,2}_{1;2,2}\sum_{i=x_{1},x_{2},x_{3}}\hat{P}_{i,x_{1}}(\Phi_{0,1}(x_{2},x_{3})Z(x_{1})-\Phi_{0,-1}(x_{2},x_{3})Z^{*}(x_{1}))+\Phi_{0,1,-1}(x_{1},x_{2},x_{3})]C^{\dagger}_{1}(x_{1})C^{\dagger}_{2}(y)C^{\dagger}_{2}(x_{3})\nonumber\\&&+\Phi_{0,1,-1}(x_{1},x_{2},x_{3})C^{\dagger}_{2}(x_{1})C^{\dagger}_{2}(x_{2})C^{\dagger}_{2}(x_{3})]|0>. 
\label{eqnarray}
\end{eqnarray}
  
This state is  expressed in terms of   the $Slater$ determinants for two and three particles $\Phi_{0,\pm1}(x_{1},x_{2})$ , $\Phi_{0,1,-1}(x_{1},x_{2},x_{3})$. ($\hat{P}_{i,x_{3}}$ is the coordinates interchange operator defined   by: 
 $\hat{P}_{i,z}F(x_{1},x_{2};x_{3}) F(x_{1},x_{2};x_{3})=\delta_{i,z}F(x_{1},x_{2};x_{3})+ \delta_{i,x_{1}} F(x_{3},y_{2};x_{1})+\delta_{i,x_{2}} F(x_{1},x_{3};x_{2})$). The three particles states can be rewritten as an  antisymmetric tensor product of  the three  single  particles states,  which obey Eq. (11):
 
$|0,1,-1;N=3>=\sum_{P}(-1)^P|0_{P(1)};N=1>|1_{P(2)};N=1>|-1_{P(3)};N=1>$.

\vspace{0.1 in}

\noindent\textbf{D-The four particles state}

\vspace{0.1 in}

The wave function for  $four$ particles  has the  structure $|n,-n,m,-m;N=4> $  with $n\neq m$. The ground state  is given by: $|1,-1,2,-2;N=4> $ with the eigenvalue $E_{g}(1,-1,2,-2;N=4)$.
From Eq. (8) we find: 
$H|1,-1,2,-2;N=4>=E(4)|1,-1,2,-2;N=4>$, $\eta|1,-1,2,-2;N=4>=0 $, $E|1,-1,2,-2;N=4>=0$  and $\gamma |1,-1,2,-2;N=4>=0$  we obtain a set of equations for the spinor components   $f_{1111}(x_{1},x_{2},x_{3},x_{4})$, $f_{1112}(x_{1},x_{2},x_{3},x_{4})$,$f_{1122}(x_{1},x_{2},x_{3},x_{4})$ ,$f_{1222}(x_{1},x_{2},x_{3},x_{4})$ and $f_{2222}(x_{1},x_{2},x_{3},x_{4})$: 

$4f_{1111}(x_{1},x_{2},x_{3},0)=\hspace{0.05 in}f_{1112}(x_{1},x_{2},x_{3},0)$\hspace{0.1 in};

 $[4\partial_{x_{4}}f_{1111}(x_{1},x_{2},x_{3},x_{4})+\partial_{x_{4}}f_{1112}(x_{1},x_{2},x_{3},x_{4}))]_{x_{4}=0}=0$, 

$4f_{2222}(x_{1},x_{2},x_{3},0)= -f_{1222}(0,x_{1},x_{2},x_{3})$; 

 $[4\partial_{x_{4}}f_{2222}(x_{1},x_{2},x_{3},x_{4})-\partial_{x_{4}}f_{1222}(x_{1},x_{2},x_{3},x_{4})]_{x_{4}=0}=0$, 

$3f_{1112}(x_{1},x_{2},0,x_{4})=-2 f_{1122}(x_{1},x_{2},x_{3},0)$;

$[3\partial_{x_{4}}f_{1112}(x_{1},x_{2},x_{3},x_{4})-2\partial_{x_{4}}f_{1122}(x_{1},x_{2},x_{3},x_{4})]_{x_{4}=0}=0$, 

$3f_{1222}(x_{1},x_{2},x_{3},0)=-2 f_{1221}(x_{1},x_{2},x_{3},0)$;

$[3\partial_{x_{4}}f_{1222}(x_{1},x_{2},x_{3},x_{4})+2\partial_{x_{4}}f_{1221}(x_{1},x_{2},x_{3},x_{4})]_{x_{4}=0}=0$. 

\noindent The eigenvalue and  the eigenfunction are:

$E_{g}(1,-1,2,-2;N=4)=\frac{\hbar^{2}}{2m}(\frac{2\pi}{L})^2 [(1-\hat{\varphi})^2+(-1-\hat{\varphi})^2+(2-\hat{\varphi})^2+(-2-\hat{\varphi})^2]$,  
\begin{eqnarray}
&&|1,-1,2,-2;N=4>=
\nonumber\\&&\int_{0}^{L}\,dx_{1} \int_{0}^{L}\,dx_{2}\int_{0}^{L}\,dx_{3}\int_{0}^{L}\,dx_{4}[\Phi_{1,-1,2,-2}(x_{1},x_{2},x_{3},x_{4})C^{\dagger}_{1}(x_{1})C^{\dagger}_{1}(x_{2})C^{\dagger}_{1}(x_{3})C^{\dagger}_{1}(x_{4})\nonumber\\&&+ 4\epsilon^{3,1}_{1,1,1;2}[\sum_{i=x_{1},x_{2},x_{3},x_{4}}\hat{P}_{i,x_{4}}[\Phi_{2,1,-1}(x_{1},x_{2},x_{3})(Z(x_{4}))^2-\Phi_{-2,1,-1}(x_{1},x_{2},x_{3})(Z^{*}(x_{3}))^2\nonumber\\&&+\Phi_{1,2,-2}(x_{1},x_{2},x_{3})Z(x_{4})-\Phi_{-1,2,-2}(x_{1},x_{2},x_{3})Z^{*}(x_{4})]]C^{\dagger}_{1}(x_{1})C^{\dagger}_{1}(x_{2})C^{\dagger}_{1}(x_{3})C^{\dagger}_{2}(w)\nonumber\\&&+6\epsilon^{2,2}_{1,1;2,2}[[\sum_{i=x_{1},x_{2},x_{3}}\hat{P}_{i,x_{3}}+ \sum_{i=x_{1},x_{2},x_{4}}\hat{P}_{i,x_{4}}][\Phi_{1,-1}(x_{1},x_{2})\Phi_{2,-2}(x_{3},x_{4})\nonumber\\&&+\Phi_{1,2}(x_{1},x_{2})\Phi_{-1,-2}(x_{3},x_{4})]]C^{\dagger}_{1}(x_{1})C^{\dagger}_{1}(x_{2})C^{\dagger}_{2}(x_{3})C^{\dagger}_{2}(x_{4})\nonumber\\&&+4\epsilon^{1,3}_{1;2,2,2}[\sum_{i=x_{1},x_{2},x_{3},x_{4}}\hat{P}_{i,x_{1}}[\Phi_{2,1,-1}(x_{2},x_{3},x_{4})(Z(x_{1}))^2-\Phi_{-2,1,-1}(x_{2},x_{3},x_{4})(Z^{*}(x_{1}))^2\nonumber\\&&+\Phi_{1,2,-2}(x_{2},x_{3},x_{4})Z(x_{1})-\Phi_{-1,2,-2}(x_{2},x_{3},x_{4})Z^{*}(x_{1})]]C^{\dagger}_{1}(x_{1})C^{\dagger}_{2}(x_{2})C^{\dagger}_{2}(x_{3})C^{\dagger}_{2}(x_{4})\nonumber\\&&+\Phi_{1,-1,2,-2}(x_{1},x_{2},x_{3},x_{4})C^{\dagger}_{2}(x_{1})C^{\dagger}_{2}(x_{2})C^{\dagger}_{2}(x_{3})C^{\dagger}_{2}(x_{4})]|0>  \nonumber\\&&\equiv\sum_{P}(-1)^P|1_{P(1)};N=1>|-1_{P(2)};N=1>|2_{P(3)};N=1>|-2_{P(4)};N=1>. 
\label{eqnarray}
\end{eqnarray}


Where $\Phi_{1,-1,2,-2}(x_{1},x_{2},x_{3},x_{4})$ , $\Phi_{\pm2,1,-1}(x_{1},x_{2},x_{3})$ and   $\Phi_{n,m}(x_{1},x_{2})$ are  the $Slater$ $determinant$ for $2,3$ and $4$  particles.  Here $\epsilon^{3,1}_{1,1,1;2}$ and  $\epsilon^{2,2}_{1,1;2,2}$ are the antisymmetric tensors for the ring index. 


\vspace{0.1 in}

\noindent\textbf{E- The $2M$   particles state}
 
 \vspace{0.1 in}
 
 The $2M$   particles state  is built from the  single particles states  $n_{1},..n_{k},..n_{M}$ given by Eq. (11) with vanishing total  momentum: 
\begin{eqnarray} 
&&|n_{1},-n_{2},..n_{2k-1},-n_{2k},..n_{2M-1},-n_{2M};N=2M>=\nonumber\\&& \sum_{P}(-1)^P|n_{P(1)};N=1>|-n_{P(2)};N=1>...|n_{P(2M-1)};N=1>|-n_{P(2M)};N=1>. \nonumber\\ 
\label{eqnarray}
\end{eqnarray}
The ground state  and the ground state energy are: $|1,-1,...M,-M;N=2M>_{g}=\sum_{P}(-1)^P|1_{P(1)};N=1>|-1_{P(2)};N=1>|2_{P(3)};N=1>|-2_{P(4)};N=1>....|k_{P(2k-1);N=1}>|-k_{P(2k)};N=1>...|M_{P(2M-1)};N=1>|-M_{P(2M)};N=1>$ \hspace{0.1 in};

$E_{g}(1,-1,..,k,-k,...M,-M)=\frac{\hbar^{2}}{2m}(\frac{2\pi}{L})^2\sum_{k=1}^{M} [(k-\hat{\varphi})^2+ (-k-\hat{\varphi})^2]$.



\vspace{0.4 in}

\noindent\textbf{F- The Current for equal fluxes }

\vspace{0.1 in} 

The current for equal fluxes with   $1,2,3,4$ and $2M$ particles is the same in both rings:
\begin{eqnarray}
&&J^{N=1}_{1}=\frac{<N=1;n| \hat{J}_{1}(x)|n;N=1>}{<N=1;n|n;N=1>}=[\frac{\hbar}{m}\frac{2\pi}{L}][\frac{\hat{\varphi}-n}{2L}]\hspace{0.1 in}; n=0,\pm1,\pm2..\nonumber\\&&
J^{N=1}_{1}(\hat{\varphi}=\frac{1}{2})=\frac{<N=1; \hat{\varphi}=\frac{1}{2},\chi(n)| \hat{J}_{1}(x) |\chi(n),\hat{\varphi}=\frac{1}{2};N=1>}{<N=1; \hat{\varphi}=\frac{1}{2},\chi(n) |\chi(n),\hat{\varphi}=\frac{1}{2};N=1>}= [\frac{\hbar}{m}\frac{2\pi}{L}][|\alpha_{+}|^2-|\alpha_{-}|^2][\frac{\hat{\varphi}-n}{2L}]\hspace{0.1 in}, \nonumber\\&&
J^{N=2}_{1}=\frac{<N=2;-1,1| \hat{J}_{1}(x)|1,-1;N=2>}{<N=2;-1,1|1,-1;N=2>}
=[\frac{\hbar}{m}\frac{2\pi}{L}][\frac{2\hat{\varphi}}{2L}]\hspace{0.1 in}, \nonumber\\&& 
J^{N=3}_{1}=\frac{<N=3;-1,1,0| \hat{J}_{1}(x)|0,,1,-1;N=3>}{<N=3;-1,1,0|0,,1,-1;N=3>}=[\frac{\hbar}{m}\frac{2\pi}{L}][\frac{3\hat{\varphi}}{2L}]\hspace{0.1 in}, \nonumber\\&&
J^{N=4}_{1}=\frac{<N=4;-2,2,-1,1| \hat{J}_{1}(x)|1,-1,2,-2;N=4>}{<N=4;-2,2,-1,1|1,-1,2,-2;N=4>}=[\frac{\hbar}{m}\frac{2\pi}{L}][\frac{4\hat{\varphi}}{2L}]\hspace{0.1 in}, \nonumber\\&&
J^{N=2M}_{1}=\frac{<N=2M;-M,M,...-1,1| \hat{J}_{1}(x)|1,-1,...M,-M;N=2M>_{g}}{<N=2M;-M,M,...-1,1|1,-1,...M,-M;N=2M>_{g}} =[\frac{\hbar}{m}\frac{2\pi}{L}][\frac{2M\hat{\varphi}}{2L}] . 
\label{eqnarray}
\end{eqnarray} 

The $magnetization$ $M^{(N)}$ is given  by the  $current$ $area$ product: $M^{(N)}= 2 J^{N}_{1}\frac{L^2}{4\pi}$.     
For an even number of electrons we find that the current in a single ring is twice the current in a double ring $J^{N=2M}_{single-ring}=2J^{N=2M}_{1}$. The factor of $\frac{1}{2}$ is a result  of the  two component  $spinor$  state  renormalization. 
At finite temperatures  the two rings excited  states have the form : $|1,-1,...M+p,-(M+p);N=2M>_{e}$ where $p$ are integers. This state carries the same current as  the ground state  $|1,-1,...M,-M;N=2M>_{g}$. Therefore, we conclude that for an $even$ (fixed) number of electrons the current will be the same at any temperature!
(When the  total number of electrons fluctuates,  $N\rightarrow N \pm2$ thermal effects will  decrease   the current.)
 The situation for the $odd$ number of electrons is  different. Even  for the   two  states  $|1,-1,...M,-M,n=(M+p);N=2M+1>$  and $|1,-1,...M,-M,n=-(M+p);N=2M+1>$ we   have  different eigenvalues and   at finite temperatures these  states  carry a   different current. 
 Therefore, the total current carried by all the states will be  reduced like we have for  a single  ring  where the unrestricted structure of the wave function  allows any configuration of momenta, which  generate an antisymmetric wave function in space:  $f^{(single-ring)}(x_{1},x_{2},...x_{N=2M})=\Phi_{n_{1},n_{2},....n_{2M}}(x_{1},x_{2},...x_{N=2M})$. 
To probe this $even$-$odd$ structure we propose to attach  a gate voltage to the rings. As a result, the magnetization   will vary  with the varying  gate voltage.   


\vspace{0.1 in} 
 
\section{The wave function for  opposite fluxes}

\vspace{0.1 in} 

For this case the constraint operator $\gamma$  is  modified to:  $\gamma=[(-i\partial_{x}-\frac{2\pi}{L}\hat{\varphi})(C_{1}(x)+C_{2}(x))]|_{x=0,L}$. For the single particle case  we find  the following  boundary conditions:

 $ f_{1}(x=0)=f_{2}(x=0)$ \hspace{0.1 in};
  $-i[\partial_{x}f_{1}(x)+\partial_{x}f_{2}(x)]|_{x=0,L}=\frac{2\pi}{L}\hat{\varphi}[f_{1}(x)+f_{2}(x)]|_{x=0,L}$.
  
\noindent We find that for this case the wave function must vanish. Only for integer values of flux  $n=integer= \hat{\varphi}$  we have finite  solutions  $f_{1}(x)= f_{2}(x)=e^{i\frac{2\pi n}{L}x}$ with  a vanishing persistent current.
  This result is in agreement with the fact  that at the common point between the rings the fluxes  must satisfy $\hat{\varphi}_{2}= \hat{\varphi}_{1}+ n$.   Therefore, the  boundary condition can be satisfied  for this case only if the the wave function vanishes  at the common point . 
We mention that for two separated rings threated by opposite fluxes the magnetization will be zero only at  the symmetry points.  This result allows to control the current in one ring by reversing the flux in the second ring. 

\section{Two coupled Cylinders}

\vspace{ 0.1 in} 

In order to build a theory which can be compared with the experiment we have to consider effects of interactions and effects of finite width geometry [13].
For realistic considerations the point contact  between two rings is replaced   by two  $narrow$ $cylinders$ of height $d\ll L$ which are in contact at the point  $(x=0,0\leq z\leq d$ 
  The $gluing$ condition is implemented by two $narrow$ $cylinders$ of height  replacing  the constraints  in Eq. (6) by $\eta(z,x=0)|\chi,N>=0$ and $\gamma(z,x=0)|\chi,N>=0$.  In the absence of disorder we obtain  for each transversal channel  $r=1,2..r_{max}$ one dimensional constraints: 
$\eta_{r}(x=0)|\chi,N>=0;$ \hspace{0.2 in} $\gamma_{r}(x=0)|\chi,N>=0$.

Therefore, the current in the channel $r$ is the same as the result given in Eq. (17).
For $N$ electrons the current will be determined by the partition of $N$  electrons in the different channels :
$N= N_{1}+N_{2}+..N_{r}...+N_{r_{max}}$

In the absence of disorder the current in cylinder  one, at T=0 will be given by:

$J^{N}_{1} =[\frac{\hbar}{m}\frac{2\pi}{L}][\frac{2(N_{1}+N_{2}+..N_{r}...N_{r_{max}})\hat{\varphi}}{2L}] \equiv [\frac{\hbar}{m}\frac{2\pi}{L}][\frac{2N\hat{\varphi}}{2L}]$.


  
  

\vspace{0.1 in}

\section{conclusion}

To conclude, a new method for applying    constraints   has been presented. This method  has been used to compute the wave function for coupled rings.  
For an  even number of electrons, only states with total vanishing momentum are allowed giving rise to a large  persistent current and magnetization.  For odd number of electrons at finite temperature the current and the magnetization are suppressed. We propose to confirm this even-odd  effect by attaching the two rings to a varying gate voltage. Reversing the flux  in one  ring will cause the current to vanish in both  rings. 
We construct the many article ground state   which obey  the constraints and show that not all the many particle states which are build from single particle states  which obey the constraints  are allowed. 

\vspace{ 0.1 in}  




\vspace{ 0.1 in}  

\begin{bibliography}{99}

\noindent 1. K. Sasaki, Y. Kewazoe, and R. Saito, Physics Letters A {\bf 321}, 369-375 (2005).\\
2. H. Aoki, J. Phys. C. {\bf 18}, 1885-1890 (1981).\\
3. Paul A. M. Dirac, ``Lectures on Quantum Mechanics,'' Belfer Graduate School of Science,     Yeshiva University, New York, 1964.\\
4. M. Henneaux and Claudio Teitelboim "Quantization of Gauge Systems", Princeton University Press (1992).\\
5. Y. Avishai and M. Luck, J. Phys. A: Math. Theory {\bf 42}, 175301 (2009).  
\\
6. S. Itoh and S. Ihara Phys. Rev. B. {\bf 48}, 8323 (1993).\\
7. H. Bluhm, N.C. Koshnick, J.A. Bert, M.E. Huber, and K.A. Moler. Cond-Mat/08104\\
8. Y. Aharonov and D.Bohm, Phys. Rev. {\bf 115}, 485 (1959). \\
9.  N. Byers and C.N. Yang, Phys. Rev. Lett. {\bf 7}, 46 (1961). \\
10.  M.Buttiker, Y. Imry, and R. Landauer, Phys. Lett. A {\bf 96}, 365 (1983).\\
11. L.P. Levy, G. Dolan, J. Dunsmuir, and H. Bouchiat, Phys. Rev. Lett. {\bf 64}, 2074 (1990).\\  
12. D. Schmeltzer, J. Phys. Condens. Matter  {\bf 20}, 335205 (2008). 

13.M.Grochol et al. Phys.Rev.B 74, 115416 (2006)
\end{bibliography}

\end{document}